\documentclass[12pt]{iopart}
\usepackage{amssymb}
\usepackage{graphicx}
\DeclareMathAlphabet{\mathpzc}{OT1}{pzc}{m}{it}

\begin{document}

\title[]{Spatial confinement effects on quantum field theory using nonlinear coherent states approach}

\author{M. Bagheri Harouni, R. Roknizadeh, M. H. Naderi}

\address{Quantum Optics Group,
Physics Department, University of Isfahan}
\ead{\mailto{m-baghreri@phys.ui.ac.ir},\
\mailto{rokni@sci.ui.ac.ir},\
\mailto{mhnaderi2001@yahoo.com}}
\begin{abstract}
We study some basic quantum confinement effects through
investigation a deformed harmonic oscillator algebra. We show
that spatial confinement effects on a quantum harmonic oscillator
can be represented by a deformation function within the framework
of nonlinear coherent states theory. Using the deformed algebra,
we construct a quantum field theory in confined space. In
particular, we find that the confinement influences on some
physical properties of the electromagnetic field and it
 gives rise to nonlinear interaction. Furthermore, we propose a
 physical scheme to generate the nonlinear coherent states
 associated with the electromagnetic field in a confined region.
\end{abstract}

\maketitle

\section{Introduction}
The physical size and shape of the materials strongly effect
 the nature, the dynamics of the electronic excitations, the lattice vibrations,
 and the dynamics of carriers. For example, in the mesoscopic systems, the
dimension of system is comparable with the coherence length of
carriers and this leads to some new phenomena that they do not
appear in a bulk semiconductor, such as quantum interference
between carrier's motion \cite{mesos}. In these physical systems
different particles are confined in a small space and interact
with each other. As usual, we use quantum field theory (QFT) and
second quantization procedure for considering interacting many
particles physical systems. Standard QFT is based on quantum
mechanics on an infinite line without any boundaries. However, the
presence of infinite walls in standard QFT can detect vacuum
effect of electromagnetic field and gives rise to Casimir effect
\cite{casimir}. Hence, in a system with small dimensions we
expect some new phenomena appear, and barriers effects show themselves. \\
Recent progress in growth techniques and development of
micromachinig technology in designing mesoscopic systems and
nanostructures, have led to intensive theoretical \cite{theory}
and experimental investigations \cite{expri} on electronic and
optical properties of those systems. The most important point
about the nanoscale structures is that the quantum confinement
effects play the center-stone role. One can even say in general
that recent success in nanofabrication technique has resulted in
great interest in various artificial physical systems with usual
phenomena driven by the quantum confinement (quantum dots,
quantum wires and quantum wells). A number of recent experiments
have demonstrated that isolated semiconductor quantum dots are
capable
 of emitting light \cite{1}. It becomes possible to combine high-Q
 optical microcavities with quantum dot emitters as the active
 medium \cite{2}. Furthermore, there are many theoretical
 attempts for understanding the optical and electronic properties of
 nanostructures especially semiconductor quantum dots \cite{3}. Because of
  intensive researches in this area, it is reasonable to
 consider the finite size effects on the EM field including the quantization of the EM
 field in confined regions that their sizes are of order of
 electromagnetic wavelength, such as microcavities. On the other hand, a nanostructure
 such as quantum dot, is a system that carrier's motion is
 confined inside a small region, and during the interaction with other
 systems, the generated excitations such as phonons, excitons,
 plasmons are confined in small region. Hence we want to answer this question: what
 are the spatial confinement effects on excitation states in quantum
 field theoretical description of nanostructures? It seems that
 to answer this question we need to know the confinement
 and boundary conditions effects in QFT. First, we consider
 spatial confinement effect on a simple quantum harmonic oscillator and then
 we shall use this oscillator in quantizing the fields.\\ \indent
 As mentioned before, the standard QFT is based on the quantum mechanics
 on an infinite line. In the canonical QFT the main tool is quantum
 oscillator. Energy eigenvalues of quantum harmonic oscillator
 (QHO) is given by $E_n=(n+\frac{1}{2})\hbar\omega$, and these
 successive energy levels were interpreted as being obtained by
 creation of a quantum particle of energy $\hbar\omega$. This
 interpretation of the energy spectrum of QHO was successfully
 used in the second quantization formalism \cite{field}. Plank's
 hypothesis is realized in the second quantization formalism by using
 creation and annihilation operators of the QHO. This realization
 is obtained for QHO defined on an infinite line.\\ \indent
 It is reasonable to claim that, in
 considering QFT in a finite region one can use energy levels of a
 QHO confined in that finite space and therefore analyze the
 consequences of this assumption in construction of such QFT on a
 compact manifold. As we shall see in subsequent sections, the spatial confinement
 of the QHO leads to a deformed Heisenberg algebra for the ordinary
 harmonic oscillator. A
 deformed algebra is a nontrivial generalization of a given
 algebra through the introduction of one or more complex
 parameters, such that, in a certain limit of parameters the
 non-deformed
 algebra is recovered; these parameters are called
 deformation parameters. There have been several attempts to
 generalize Heisenberg algebra, and a particular deformation of
 Heisenberg algebra has led to the notion f-oscillator \cite{man1}. An
 f-oscillator is a non-harmonic system, that from mathematical
 point of view its dynamical variables (creation and
 annihilation operators) constructed from a non canonical
 transformation through
 \begin{equation}\label{defo}
\hat{A}=\hat{a}f(\hat{n})\hspace
{1cm},\hspace{1cm}\hat{A}^{\dag}=f(\hat{n})\hat{a}^{\dag},
 \end{equation}
where $\hat{a}$ and $\hat{a}^{\dag}$ are corresponding harmonic
oscillator operators and $\hat{n}=\hat{a}^{\dag}\hat{a}$. The
function $f(\hat{n})$ is called deformation function that depends
on the number of quanta and some physical parameters. The presence
of operator-valued deformation function causes the Heisenberg
algebra of the standard QHO to transform into a deformed
Heisenberg algebra. The nonlinearity in f-oscillators means
dependence of the frequency on the intensity \cite{man2}. On the
other hand, in contrast to the standard QHO, f-oscillators have
not equal spaced energy spectrum. If we confine a simple QHO
inside an infinite well, due to the spatial confinement, the
energy levels constitute a spectrum that is not equal spaced.
Therefore, in this case it is reasonable to expect to find a
corresponding f-oscillator. One of the most interesting features
of the QHO is the construction of coherent states, as the
eigenfunction of annihilation operator. As is well known
\cite{man1} one can introduce Nonlinear coherent states or
f-coherent states as the right-hand eigenstates of deformed
annihilation operator $\hat{A}$. It has been shown that these
families of generalized coherent states exhibit various
non-classical properties \cite{hame}. Due to these properties and
their applications, generation of these states is a very
important issue in the context of quantum optics. The f-coherent
states may appear as stationary states of the center-of-mass
motion of a trapped ion \cite{vogel}. Furthermore, a theoretical
scheme for generation of these states in micromaser in the frame
work of intensity-dependent Jaynes-Cummings model has been
proposed \cite{naderi}.\\ \indent It has also been shown
\cite{nondet} that there is a close connection between the
deformation function appeared in the nonlinear coherent states
algebraic structure and the non-commutative geometry of the
configuration space. Furthermore, it has been shown recently
\cite{mahdi}, that if a two-mode QHO  confined on the surface of
a sphere, can be interpreted as a single mode deformed
oscillator, whose and its quantum statistics depends on the
curvature of sphere.\\ \indent Motivated by the above-mentioned
results, in the present contribution we are intended to
investigate the spatial confinement effects on physical properties
of a standard QHO. It will be seen that the confinement leads to
deformation of standard QHO. Then we use this confined oscillator
to considering boundary effects in QFT. In a recent work
\cite{bezze} the authors have considered boundary effects in QFT
and for this purpose they have used a QHO defined on a circle and
its associated algebra, which is a realization of a deformed
Heisenberg algebra has been introduced in Ref.\cite{cur}. To
construct QFT they have used this special deformed algebra and
the calculus on a lattice without any definite commutation
relation between field operators. In this paper, we consider a
QHO confined in a one-dimensional infinite well without periodic
 boundary conditions, and we find its energy levels, as well as associated ladder
 operators. We show that the ladder operators can be interpreted as a special kind
 of the so-called f-deformed creation and annihilation operators \cite{man1}. Then, we use
 this oscillator as a basis for the canonical quantization of the electromagnetic (EM) field in a confined space.
  In Ref. \cite{swamy}
 the quantization of the
 electromagnetic field is performed by making use of the q-deformed oscillator without any
 quantization postulate. In our quantization scheme we use
 the quantization postulate and impose canonical commutation relation
 on Hamiltonian of the system under consideration. In order to keep commutation
 relation between field and its conjugate momentum we
 deform Hilbert space of the system.\\ \indent
 This paper is organized as follow: In Section 2, we review some
 physical properties of f-oscillator and its coherent states.
  In section 3 we consider the spatially confined QHO in a one-dimensional infinite well and construct
  its associated coherent states. We shall also examine some of their quantum statistical properties,
  including sub-Poissonian statistics and quadrature squeezing.
   In section 4  we use the confined oscillator under consideration and its algebra to construct
 a quantum theory of fields, and as an example we quantize
 the electromagnetic field. In Section 5 we propose a dynamical scheme for
 generating the nonlinear coherent state associated with the EM field in a confined region. Finally we summarize our
 conclusions in section 6.
 \section{f-oscillator and nonlinear coherent states}
 In this section, we review the basics of the f-deformed quantum
 oscillator and the associated coherent states known in the
 literature as nonlinear coherent states. For this purpose,
  we consider an eigenvalue problem for a given quantum physical system and we focus
 our attention on the properties of
 creation and annihilation operators, that allows to make
 transition
 between the states of discrete spectrum of the system Hamiltonian.
 As usual, we expand the Hamiltonian in its eigenvectors
 \begin{equation}
\hat{H}=\sum_{i=0}^{N-1}E_i|i\rangle\langle i|\:,
\end{equation}
where we choose $E_0=0$. We introduce the creation (raising) and
annihilation (lowering) operators as follows
\begin{equation}
\hat{A}^{\dag}=\sum_{i=}^{N-1}\sqrt{E_{i+1}}|i+1\rangle\langle i|
\hspace{1cm},\hspace{1cm}\hat{A}=\sum_{i=0}^{N-1}\sqrt{E_{i}}|i-1\rangle\langle
i|\:,
\end{equation}
so that $\hat{A}^{\dag}|N\rangle =\hat{A}|0\rangle=0$. These
ladder operators satisfy the following commutation relation
\begin{equation}
[\hat{A},\hat{A}^{\dag}]=\sum_{i=1}^{N}(E_{i+1}-E_{i})|i\rangle\langle
i|\:.
\end{equation}
Obviously if the energy spectrum is equally spaced ,because of
this condition, energy spectrum must be linear in quantum
numbers, (as in the case of ordinary QHO), then
$E_{i+1}-E_{i}=c$, where $c$ is a constant and the commutator of
$\hat{A}$ and $\hat{A}^{\dag}$ becomes a constant (a rescaled
Weyl-Heisenberg algebra). On the other hand, if the energy
spectrum is not equally spaced, the ladder operators of the
system satisfy a deformed Heisenberg algebra, i.e. their
commutator depends on quantum numbers that appear in energy
spectrum. This is one of the most
important properties of the quantum f-oscillators \cite{man1}.\\
\indent An f-oscillator is a non-harmonic system characterized by
a Hamiltonian of the harmonic oscillator form
\begin{equation}\label{hami}
\hat{H}_D=\frac{1}{2}\Omega(\hat{A}\hat{A}^{\dag}+\hat{A}^{\dag}\hat{A})\hspace{1cm}
(\hbar=1)\:,
\end{equation}
with a specific frequency $\Omega$ and deformed boson creation
and annihilation operators defined in (\ref{defo}). The deformed
operators obey the commutation relation
\begin{equation}\label{comut}
[\hat{A}\:,\:\hat{A}^{\dag}]=(\hat{n}+1)f^2(\hat{n}+1)-\hat{n}f^2(\hat{n})\:.
\end{equation}
 The f-deformed Hamiltonian
$\hat{H}_D$ is diagonal on the eingenstates $|n\rangle$ in the
Fock space and its eigenvalues are
\begin{equation}\label{energy}
E_n=\frac{\Omega}{2}[(n+1)f^2(n+1)+nf^2(n)].
\end{equation}
In the limit $f\rightarrow 1$, the ordinary expression
$E_n=\hbar\Omega (n+\frac{1}{2})$ and the usual (non-deformed)
commutation relation $[\hat{a}\:,\:\hat{a}^{\dag}]=1$ are recovered. \\
\indent Furthermore, by using the Heisenberg equation of motion
with Hamiltonian (\ref{hami}) we have
\begin{equation}
i\frac{d\hat{A}}{dt}=[\hat{A}\:,\:\hat{H}_D]\hspace{1cm}(\hbar=1).
\end{equation}
We obtain the following solution to the Heisenberg equation of
motion for f-deformed operators $\hat{A}$ and $\hat{A}^{\dag}$
defined in equation (\ref{defo})
\begin{equation}
\hat{A}(t)=e^{-i\Omega
G(\hat{n})t}\hat{A}(0)\hspace{0.5cm},\hspace{0.5cm}\hat{A}^{\dag}(t)=\hat{A}^{\dag}(0)e^{i\Omega
G(\hat{n})t},
\end{equation}
where
\begin{equation}
G(\hat{n})=\frac{1}{2}\left((\hat{n}+2)f^2(\hat{n}+2)-\hat{n}f^2(\hat{n})\right).
\end{equation}
In this sense, the f-deformed oscillator can be interpreted as a
nonlinear oscillator whose frequency of vibrations depends
explicitly on its number of excitation quanta \cite{man2}. It is
interesting to point out that recent studies \cite{naderi1} have
revealed strictly physical relationship between the nonlinearity
concept resulting from f-deformation and some nonlinear optical
effects, e.g., Kerr nonlinearity, in the context of atom-field
interaction. \\ \indent The nonlinear transformation of the
creation and annihilation operators leads naturally to the notion
of nonlinear coherent states or f-coherent states. The nonlinear
coherent states $|\alpha\rangle_f$ are defined as the right-hand
eigenstates of the deformed operator $\hat{A}=\hat{a}f(\hat{n})$
\begin{equation}\label{coh}
\hat{A}|\alpha\rangle_f=\alpha|\alpha\rangle_f\:.
\end{equation}
From Eq.(\ref{coh}) one can obtain an explicit form of the
nonlinear coherent states in a number state representation
\begin{equation}
|\alpha\rangle_f=C\sum_{n=0}^{\infty}\alpha^nd_n|n\rangle,
\end{equation}
where the coefficients $d_n$'s and normalization constant $C$
 are respectively given by
\begin{eqnarray}\label{coh1}
d_0&=&1\hspace{0.5cm},\hspace{0.5cm}d_n=\left(\sqrt{n!}f(n)!\right)^{-1}\hspace{0.5cm},\hspace{0.5cm}
f(n)!=\prod_{j=1}^nf(j),\\
C&=&\left(\sum_{n=0}^{\infty} d_n^2|z|^{2n}\right)^{\frac{-1}{2}}.
\end{eqnarray}
In recent years nonlinear coherent states have been paid much
attentions because they exhibit nonclassical features \cite{hame}
and many quantum optical states, such as squeezed states, phase
states, negative binomial states and photon-added coherent states
can be viewed as a sort of nonlinear coherent states
\cite{hame1}. \\ \indent
\section{Quantum harmonic oscillator in a one dimensional infinite well}
In this section we consider a quantum harmonic oscillator
confined in a one dimensional infinite well. Many attempts have
been done for solving this problem (see \cite{CQHO}-\cite{agu},
and references therein). In most of those works, authors tried to
solve the problem numerically. But in our consideration we try to
solve the problem analytically, to reveal the relationship between
the confinement effect and given deformation function. We start
from the Schr\"{o}dinger equation (we assume $\hbar=1$)
\begin{equation}
\left[-\frac{1}{2m}\frac{d^2}{dx^2}+\frac{1}{2}kx^2+V(x)\right]\psi(x)=E\psi(x),
\end{equation}
where
\begin{displaymath}
V(x)=\left\{\begin{array}{ll} 0 & \textrm{$-a\leq x\leq a$}\\
\infty & \textrm{elsewhere}.
\end{array} \right.
\end{displaymath}
Instead of solving the Schr\"{o}dinger equation for the QHO
confined between infinite rectangular walls in positions $\pm a$,
we propose to solve the eigenvalue equation for the potential
\begin{equation}\label{pot}
V(x)=\frac{1}{2}k\left(\frac{\tan(\delta x)}{\delta}\right)^2\:,
\end{equation}
where $\delta=\frac{\pi}{2a}$, is a scaling factor depending on
the width of the well. This potential models a QHO placed in the
center of the rectangular infinite well \cite{zico}. The
potential $V(x)$ fulfills two asymptotic requirements: 1)
$V(x)\rightarrow\frac{1}{2}kx^2$ when $a\rightarrow\infty$ (free
harmonic oscillator limit). 2) $V(x)$ at equilibrium position
have the same curvature as a free QHO,
$\left[\frac{d^2V}{dx^2}\right]_{x=0}=k$. \\
Now we consider the following equation
\begin{equation}
\left[-\frac{1}{2m}\frac{d^2}{dx^2}+\frac{1}{2}k\left(\frac{\tan(\delta
x)}{\delta}\right)^2-E\right]\psi(x)=0\;.
\end{equation}
To solve analytically this equation, we use the factorization
method \cite{fac}. By changing the variable and some mathematical
manipulation, the corresponding energy eigenvalues are found as
\begin{equation}\label{eee}
E_n=\gamma'(n+\frac{1}{2})^2+\sqrt{\gamma'^2+\omega^2}(n+\frac{1}{2})+\frac{\gamma'}{4}\:,
\end{equation}
where $\gamma'=\frac{4\pi^2}{32a^2m}$,and
$\omega=\sqrt{\frac{k}{m}}$ is the frequency of the QHO. The first
term in the energy spectrum can be interpreted as the energy of a
free particle in a well, the second term denotes the energy
spectrum of the QHO, and the last term shifts energy spectrum by a
constant amount. It is evident that if $a\rightarrow\infty$ then
$\gamma'\rightarrow 0$ and the energy spectrum (\ref{eee}) reduces
to the spectrum of the free QHO. As is clear from (\ref{eee}),
different energy levels are not equally spaced, hence confining a
free QHO leads to deformation of its dynamical algebra, and we can
interpret the parameter $\gamma'$ as the deformation parameter.
In Table (\ref{malek}) the numerical results associated with the
original potential are compared with the generated results from
model potential. As is seen the results are in a good agreement
when boundary size is of order of characteristic length of the
harmonic oscillator. On the other hand, the numerical results
given in Ref. \cite{CQHO} are related to the original potential,
confined QHO in the one-dimensional infinite well. This
oscillator when approached to the boundaries of well suddenly
becomes infinite, while the model potential is smooth and
approach to infinity asymptotically. Therefore, the model
potential (\ref{pot}) is more appropriate for the physical systems
 will be considered later. \\ \indent
 If we renormalize Eq.(\ref{eee}) to energy quanta of the simple
 harmonic oscillator and introducing the new variables $n+\frac{1}{2}=l$,
$\sqrt{\frac{\gamma'^2}{\omega^2}+1}=\alpha$, and
$\gamma=\frac{\gamma'}{\omega}$ then Eq.(\ref{eee}) takes the
following form
\begin{equation}
E_l=\gamma l^2+\alpha l+\frac{\gamma}{4}.
\end{equation}
By comparing this spectrum with the energy spectrum of an
f-deformed oscillator (\ref{energy}), we find the corresponding
deformation function as
\begin{equation}\label{f1f}
f(\hat{n})=\sqrt{\gamma\hat{n}+\alpha}.
\end{equation}
This function leads to spectrum Eq.(\ref{eee}). Furthermore, the
ladder operators associated with the confined oscillator under
consideration can be written in terms of the conventional
(non-deformed) operators $\hat{a}$ , $\hat{a}^{\dag}$ as follows
\begin{equation}
\hat{A}=\hat{a}\sqrt{\gamma\hat{n}+\alpha}\hspace{1cm},
\hspace{1cm}\hat{A}^{\dag}=\sqrt{\gamma\hat{n}+\alpha}\,\,\hat{a}^{\dag}.
\end{equation}
These two operators satisfy the following commutation relation
\begin{equation}
[\hat{A},\hat{A}^{\dag}]=\gamma(2\hat{n}+1)+\alpha.
\end{equation}
It is obvious that in the limiting case $a\rightarrow\infty$
($\gamma\rightarrow 0$,$\alpha\rightarrow 1$), the right hand side
of the above commutation relation becomes independent of
$\hat{n}$, and the deformed algebra reduces to a the conventional
Weyl-Heisenberg algebra for a
free QHO.\\
Classically, harmonic oscillator is a particle that attached to an
ideal spring, and can oscillate with specific amplitude. When that
particle be confined, boundaries can affect particle's motion if
the boundaries position be in a smaller distance in comparison
with a characteristic length that particle oscillate in it. This
characteristic length for the QHO is given by
$\frac{\hbar}{m\omega}$ where $(\hbar=1)$ , and if
$2a\leq\frac{1}{m\omega}$, then the presence of boundaries affect
the behavior of QHO, otherwise it behaves like a free QHO.
Therefore, one can interpret $l_0=\frac{1}{m\omega}$ as a scale
length where the deformation effects become relevant.
\subsection{Coherent states of confined oscillator}
\begin{figure}
\begin{center}
\includegraphics[angle=0,width=.5\textwidth]{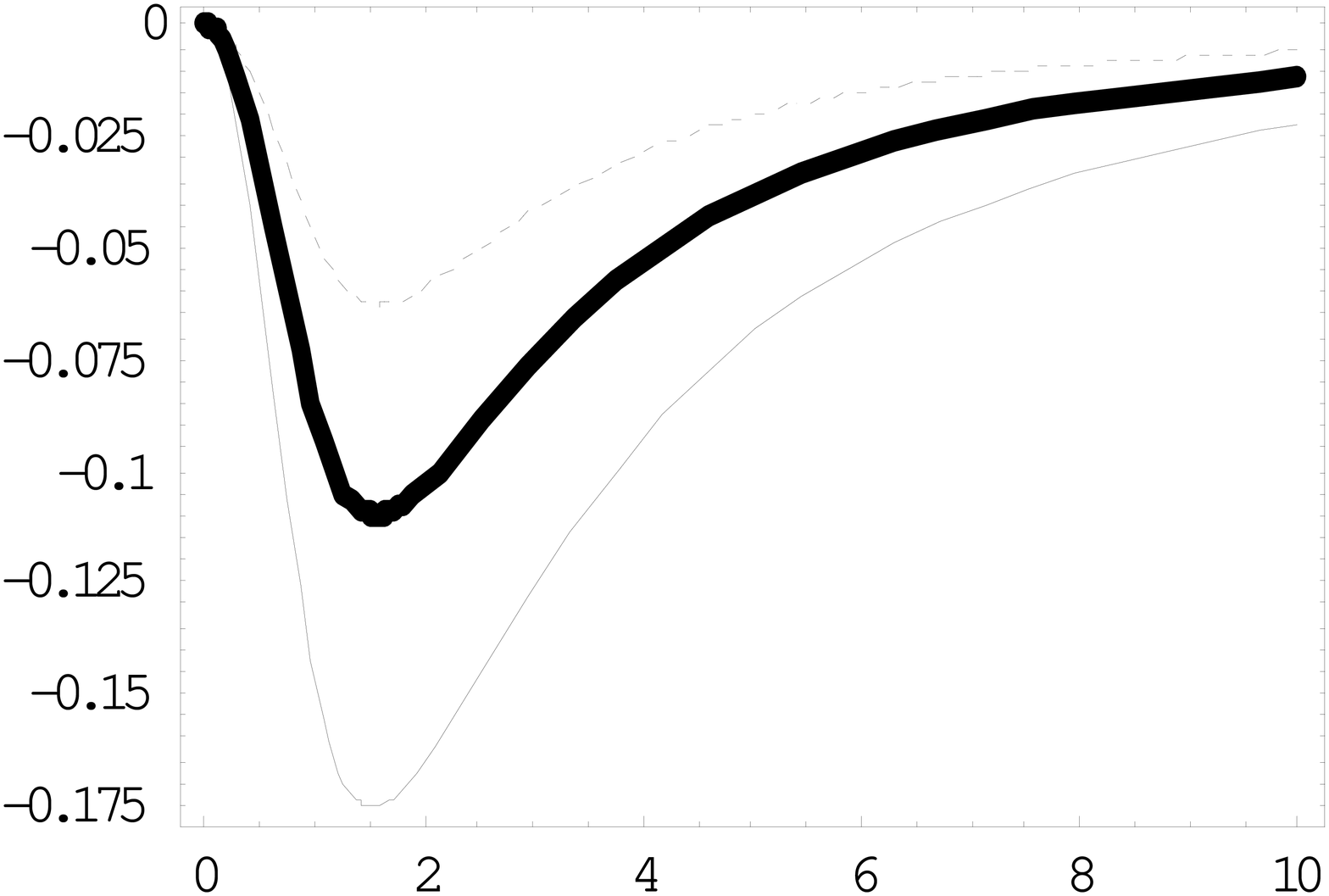}
 \caption{Plots of the Mandel parameter versus the dimensionless parameter
  $\frac{a}{l_0}$. The dotted correspond to $|\beta|^2=0.5$, the next
 correspond to $|\beta|^2=1$, and the uppest for $|\beta|^2=1.5$ } \label{f1}
\end{center}
\end{figure}

\begin{figure}
\begin{center}
\includegraphics[angle=0,width=.5\textwidth]{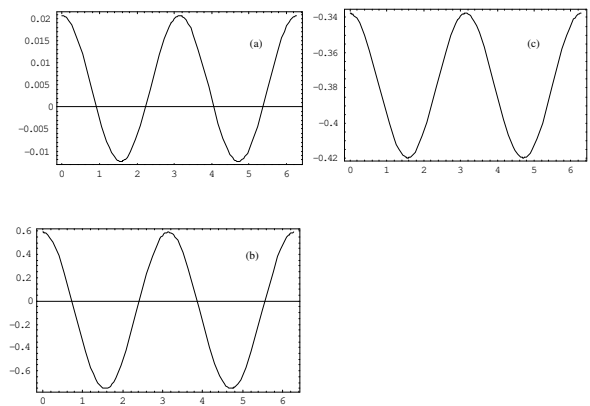}
 \caption{Plots of $s_{x_a}$ versus $\phi$ for $|\beta|^2=4$. In figure (a) we choose $a=0.5$, in figure (b) $a=1$
 , and figure (c) $a=2.5$ (these value of $a$ are renormalized to $l_0$).} \label{f2}
\end{center}
\end{figure}

\begin{figure}
\begin{center}
\includegraphics[angle=0,width=.5\textwidth]{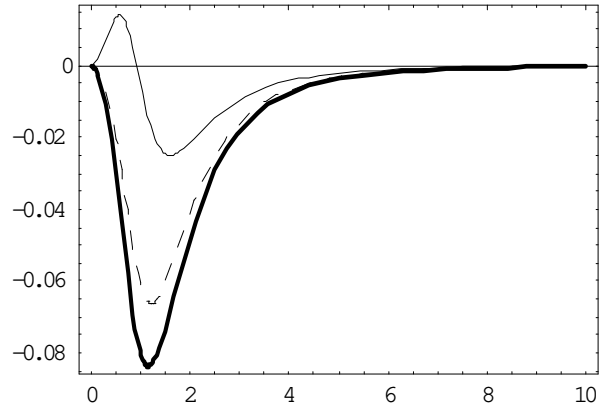}
 \caption{Plots of $s_{x_a}$ versus the dimensionless parameter
  $\frac{a}{l_0}$ for different phases and $|\beta|^2=1$. Dotted line, line and bold line
  ,respectively, correspond to $\phi=100$, $\phi=110$ and $\phi=90$.} \label{f3}
\end{center}
\end{figure}

\begin{figure}
\begin{center}
\includegraphics[angle=0,width=.5\textwidth]{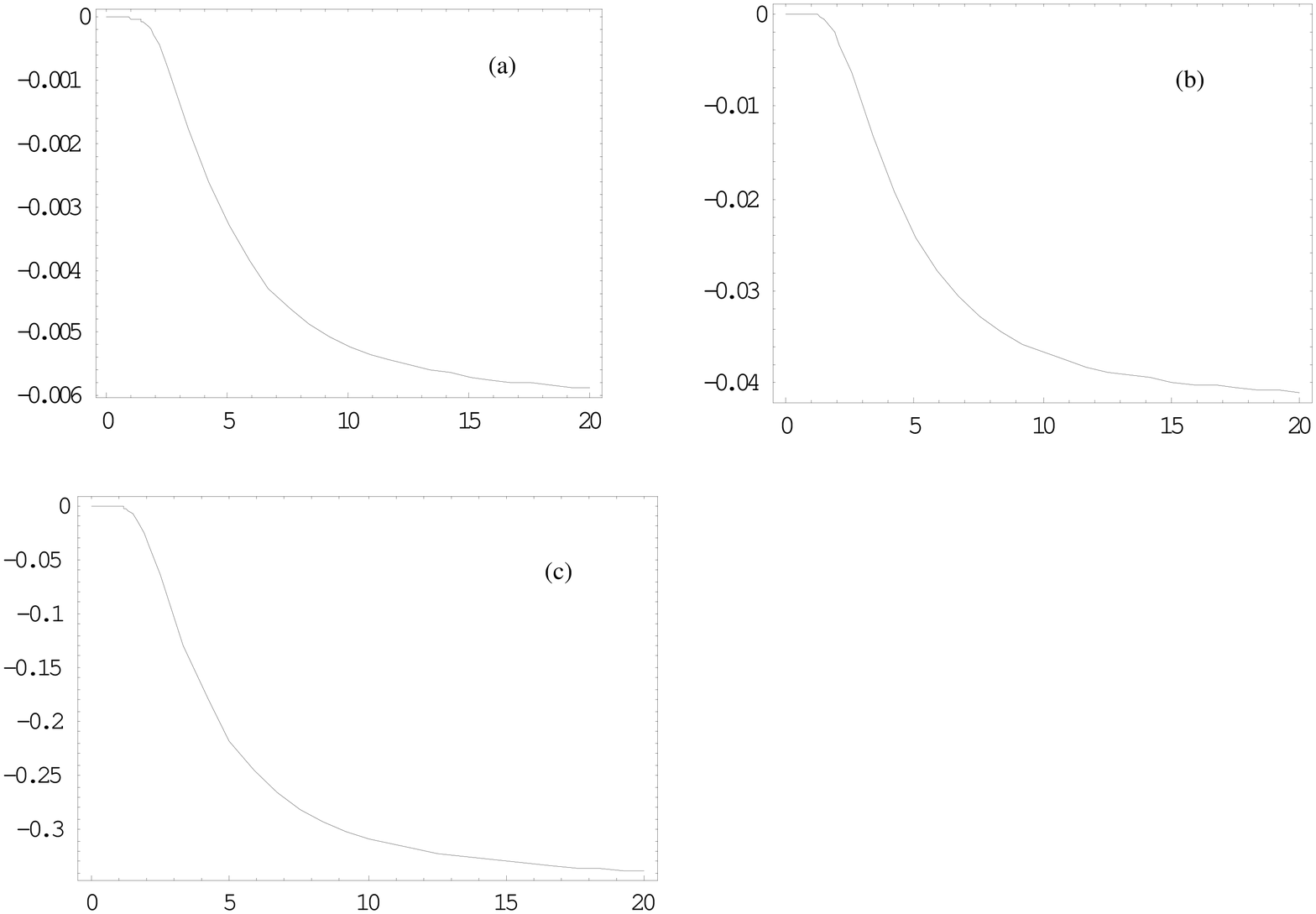}
 \caption{Plots of deformed squeezing parameter $S_{X_A}$ versus the dimensionless parameter
  $\frac{a}{l_0}$. Figures (a), (b) and (c), respectively correspond to $|\beta|^2=1$, $|\beta|^2=1.5$ and
   $|\beta|^2=2.5$.} \label{f4}
\end{center}
\end{figure}

Now, we focus our attention on the coherent states associated
with the QHO under consideration. As usual, we define coherent
states as the right-hand eigenstates of the deformed annihilation
operator
\begin{equation}\label{nlcs}
\hat{A}|\beta\rangle_f=\beta|\beta\rangle_f.
\end{equation}
From (\ref{nlcs}) we can obtain an explicit form of the state
$|\beta\rangle_f$ in a number state representation
\begin{equation}
  |\beta\rangle_f=\mathcal{N}\sum_n\frac{\beta^n}{[f(n)]!\sqrt{n!}}|n\rangle,
\end{equation}
where
$\mathcal{N}=\left(\sum_n\frac{|\beta|^2}{[f(n)!]^2n!}\right)^{-\frac{1}{2}}$
is the normalization factor, $\beta$ is a complex number, and the
deformation function $f(n)$ is given by Eq.(\ref{f1f}). The
ensemble of states $|\beta\rangle_f$ labeled by the single
complex number $\beta$ is called a set of coherent states if the
following conditions are satisfied \cite{klauder}:
\begin{itemize}
  \item normalizability
\begin{equation}\label{c1}
  _f\langle\beta|\beta\rangle_f=1,
\end{equation}
  \item continuity in the label $\beta$
\begin{equation}
  |\beta-\beta'|\rightarrow0
  \hspace{0.5cm}\Rightarrow\hspace{0.5cm}\|\;|\beta\rangle_f-|\beta'\rangle_f\|\rightarrow0,
\end{equation}
  \item resolution of the identity
\begin{equation}\label{mmm}
  \int_c
  d^2\beta|\beta\rangle_{f}{ }_f\langle\beta|w(|\beta|^2)=\hat{I},
\end{equation}
where $w(|\beta|^2)$ is a proper measure that ensures the
completeness and the integration is restricted to the part of the
complex plane where normalization converges.
\end{itemize}
The first two conditions can be proved easily. For the third
condition, we choose the normalization constant as
\begin{equation}
  \mathcal{N}^2=\frac{|\beta|^{\alpha}}{I_\alpha^\gamma(2|\beta|)},
\end{equation}
where
\begin{equation}
I_\alpha^\gamma(x)=\sum_{s=0}^\infty\frac{1}{s!(\gamma
s+\alpha)!}(\frac{x}{2})^{2s+\alpha},
\end{equation}
is similar to the Modified Bessel function of the first kind of
the order $\alpha$ with the series expansion
$I_\alpha(x)=\sum_{s=0}^\infty\frac{1}{s!(s+\alpha)!}(\frac{x}{2})^{2s+\alpha}$.
Resolution of the identity of deformed coherent states can be
written as
\begin{eqnarray}
\int
d^2\beta|\beta\rangle_f\langle\beta|w(|\beta|)=&&\pi\sum_n|n\rangle\langle
n |\frac{1}{n!(\gamma n+\alpha)!}\int_0^\infty
d|\beta||\beta||\beta|^{2n}\\ \nonumber
&&\times\frac{|\beta|^\alpha}{I_\alpha^\gamma(2|\beta|)}w(|\beta|).
\end{eqnarray}
Now we introduce the new variable $|\beta|^2=x$ and the measure
\begin{equation}
  w(\sqrt{x})=\frac{8}{\pi}I_\alpha^\gamma(2\sqrt{x})K_m(2\sqrt{x})\sqrt{x}^l,
\end{equation}
where $K_m(x)$ is the modified Bessel function of the second kind
of the order $m$, $m=(\gamma-1)n+\alpha$ and $l=(\gamma-1)n+1$.
Using the integral relation $\int_0^\infty
K_\nu(t)t^{\mu-1}dt=2^{\mu-2}\Gamma\left(\frac{\mu-\nu}{2}\right)\Gamma\left(\frac{\mu+\nu}{2}\right)$
\cite{bessel}, we obtain
\begin{equation}
\int d^2\beta|\beta\rangle_f
 { }_f\langle\beta|w(|\beta|)=\sum_n|n\rangle\langle n |=\hat{1}.
\end{equation}
 \indent We therefore conclude that the states $|\beta\rangle_f$ qualify as coherent states in the sense
described by the condition (\ref{c1})-(\ref{mmm}). We now proceed
to examine some nonclassical properties of the nonlinear coherent
states $|\beta\rangle_f$. As an important quantity, we consider
the variance of the number operator $\hat{n}$. Since for the
conventional (non-deformed) coherent states the variance of
number operator is equal to its average, deviation from
Poissonian statistics can be measured with the Mandel parameter
\cite{mandel}
\begin{equation}
  M=\frac{(\Delta
  n)^2-\langle\hat{n}\rangle}{\langle\hat{n}\rangle}.
\end{equation}
This parameter vanishes for the Poisson distribution, is positive
for super-Poissonian distribution (photon bunching effect), and
is negative for a sub-Poissonian distribution
 (photon antibunchig effect). \\
Figure \ref{f1} shows the size dependence of the Mandel parameter
for different values of dimensionless parameter$\frac{a}{l_0}$. As
is seen, the Mandel parameter exhibit sub-Poissonian
 statistics and with further increasing values of $a$ it is finally stabilized at an asymptotical zero value
 corresponding to the Poissonian statistics.\\ \indent
As another important nonclassical property we examine the
quadrature squeezing. For this purpose we first consider the
conventional quadrature operators $\hat{X}_a$ and $\hat{Y}_a$
defined in terms of undeformed operators $\hat{a}$ and
$\hat{a}^\dag$ as
\begin{equation}
  \hat{X}_{a}=\frac{1}{2}(\hat{a}e^{i\phi}+\hat{a}^{\dag}e^{-i\phi})\hspace{1cm}\hat{Y}_a=\frac{1}{2i}(\hat{a}e^{i\phi}
  -\hat{a}^{\dag}e^{-i\phi}).
\end{equation}
The commutation relation for $\hat{a}$ and $\hat{a}^{\dag}$ leads
to the following uncertainty relation
\begin{equation}
  (\Delta X_a)^2(\Delta
  Y_a)^2\geq\frac{1}{16}|\langle[\hat{X}_a,\hat{Y}_a]\rangle|^2=\frac{1}{16}.
\end{equation}
For the vacuum state $|0\rangle$, we have $(\Delta X_a)^2=(\Delta
Y_a)^2=\frac{1}{4}$ and hence $(\Delta X_a)^2(\Delta
Y_a)^2=\frac{1}{16}$. A given quantum state of the QHO is said to
be squeezed when the variance of one of the quadrature components
$\hat{X}_a$ and $\hat{Y}_a$ satisfies the relation
\begin{equation}
  (\Delta O_{a})^2<(\Delta O_{a})^2_{vacuum}=\frac{1}{4}\hspace{0.5cm}  (O_a=X_a\hspace{0.3cm}or\hspace{0.3cm}
  Y_a).
\end{equation}
The degree of quadrature squeezing can be measured by the
squeezing parameter $s_O$ defined by
\begin{equation}
  s_O=4(\Delta O_a)^2-1.
\end{equation}
Then, the condition for squeezing in the quadrature component can
be simply written as $s_O<0$. In figure \ref{f2} we have plotted
 the parameter $s_O$ corresponding to the squeezing of $\hat{X}_a$ with respect to
 the phase angle $\phi$ for three different values of $a$. This diagram shows that
the state $|\beta\rangle_f$ exhibit squeezing for different values
of the confinement size, and maximum value of squeezing occurs
when $a=1$. Figure \ref{f3} shows the plot of $s_{X_a}$ versus the
dimensionless parameter $\frac{a}{l_0}$ for different values of
phase. As is seen, with the increasing value of $\frac{a}{l_0}$
quadrature
squeezing is is stabilized to zero, according to Mandel parameter.\\
Let us also consider the deformed quadrature operators $X_A$ and
$Y_A$ defined in terms of the deformed operator $\hat{A}$ and
$\hat{A}^{\dag}$
\begin{equation}
  \hat{X}_{A}=\frac{1}{2}(\hat{A}e^{i\phi}+\hat{A}^{\dag}e^{-1\phi})\hspace{1cm}\hat{Y}_A=\frac{1}{2i}(\hat{A}e^{i\phi}
  -\hat{A}^{\dag}e^{-i\phi}).
\end{equation}
By considering the commutation relation for the deformed operators
 $\hat{A}$ and $\hat{A}^{\dag}$ (\ref{comut}), the squeezing condition
 for the deformed quadrature operators $\hat{O}_A$ can be written
 as
\begin{equation}
S=4(\Delta
O_A)^2-\langle(\hat{n}+1)f^2(\hat{n}+1)\rangle+\langle\hat{n}f^2(\hat{n})\rangle<0,
\end{equation}
where $O=X_A$ or $Y_A$. Figure \ref{f4} shows the plots of
$S_{X_A}$ versus dimensionless parameter $\frac{a}{l_0}$ for three
different values of $|\beta|^2$. As is seen, the deformed
quadrature operator always exhibits squeezing. \linespread{1}
\begin{table}\label{malek}
\begin{center}
\caption{(Calculated energy levels of the confined QHO in a one
dimensional infinite well by using our model potential in
comparison with the numerical result given in Ref.\cite{CQHO})}
\begin{tabular*}{\textwidth}{@{}l*{15}{@{\extracolsep{0pt plus12pt}}l}}
\br
state& boundary size& model potential& numerical results\\
\hline 0& a=0.5& 4.98495312& 4.95112932\\
0& 1& 1.41089325& 1.29845983\\
0& 2& 0.67745392& 0.53746120\\
0& 3& 0.57321464& 0.50039108\\
0& 4& 0.54003728& 0.50000049\\
\hline 1& a=0.5& 19.88966157& 19.77453417\\
1& 1& 5.46638033& 5.07558201\\
1& 2& 2.34078691& 1.76481643\\
1& 3& 1.85672176& 1.50608152\\
1& 4& 1.69721813& 1.50001461\\
\hline 2& a=0.5& 44.66397441& 44.45207382\\
2& 1& 11.98926850& 11.25882578\\
2& 2& 4.62097017& 3.39978824\\
2& 3& 3.41438455& 2.54112725\\
2& 4& 3.00861155& 2.50020117\\
\hline 3& a=0.5& 79.30789166& 78.99692115\\
3& 1& 20.97955777& 19.89969649\\
3& 2& 7.51800371& 5.58463907\\
3& 3& 5.24620303& 3.66421964\\
3& 4& 4.47421754& 3.50169153\\
\hline
4& a=0.5& 123.82141330& 123.41071050\\
4& 1& 32.43724814& 31.00525450\\
4& 2& 11.03188752& 8.36887442\\
4& 3& 7.35217718& 4.95418047\\
4& 4& 6.09403610& 4.50964099\\
\br
\end{tabular*}
\end{center}
\end{table}
\linespread{1.4}
\section{Quantization of the EM field in confined region}
\subsection{Mathematical preliminary}
In this section, at first we introduce a mathematical structure on
Hilbert space developed recently \cite{tavasol}. We consider an
abstract Hilbert space $\mathfrak{H}$. Let $\hat{T}$ be an
operator on this space with the properties:
\begin{itemize}
  \item $\hat{T}$ is densely defined and closed; we denote its domain by
  $\mathcal{D}(T)$.
  \item $\hat{T}^{-1}$ exists and is densely defined, with domain
  $\mathcal{D}(T^{-1})$.
  \item The vectors $\phi_n\in\mathcal{D}(T)\cap\mathcal{D}(T^{-1})$ for all $n$ and there exist non-empty open
  sets $\mathcal{D}_T$ and
$\mathcal{D}_{T^{-1}}$ in $\mathbb{C}$ such that
$\eta_z\in\mathcal{D}(T), \forall z\in\mathcal{D}_T$ and
$\eta_z\in\mathcal{D}(T^{-1}), \forall z\in\mathcal{D}^{T^{-1}}$.
\end{itemize}
Note that the first condition implies that the operator
$\hat{T}^\ast\hat{T} =\hat{F}$ is self-adjoint (here $\ast$ shows
adjoint of operators). Due to action of the operator $\hat{T}$,
the Hilbert space is transformed and orthogonal basis $\phi_n$ is
transformed to a nonorthogonal basis. This new basis can be
considered orthogonal due to a new
scalar product.\\
We define the two new Hilbert spaces:
\begin{itemize}
  \item $\mathfrak{H}_F$, which is the completion of the set $\mathcal{D}(T)$ in the scalar product
\begin{equation}\label{hehehe}
  \langle
  \psi|\phi\rangle_F=\langle\psi|\hat{T}^{\ast}\hat{T}\phi\rangle_{\mathfrak{H}}=\langle\psi|\hat{F}
  \phi\rangle_{\mathfrak{H}}.
\end{equation}
The set ${\phi_n^F=\hat{T}^{-1}\phi_n}$ is orthonormal in
$\mathcal{H}_F$ and the map $\phi\rightarrow \hat{T}^{-1}\phi,
\phi\in\mathcal{D}(T^{-1})$ extends to a unitary map between
$\mathfrak{H}$ and $\mathfrak{H}_F$ . If both $\hat{T}$ and
$\hat{T}^{-1}$ are bounded, $\mathfrak{H}_F^{-1}$ coincides with
$\mathfrak{H}$ as a set.
  \item $\mathfrak{H}_F$, which is the completion of $\mathcal{D}(T^{\ast-1})$ in the scalar product
  \begin{equation}\label{rrt}
  \langle
  \psi|\phi\rangle_F^{-1}=\langle\psi|\hat{T}^{-1}\hat{T}^{\ast-1}\phi\rangle_{\mathfrak{H}}=\langle\psi|\hat{F}^{-1}
  \phi\rangle_{\mathfrak{H}}.
\end{equation}
The set ${\phi_n^{F^{-1}}=\hat{T}\phi_N}$ is orthonormal in
$\mathfrak{H}_F^{-1}$ and the map $\phi\rightarrow \hat{T}\phi,
\phi\in\mathcal{D}(T)$ extends to a unitary map between
$\mathfrak{H}_F$ and $\mathfrak{H}_F^{-1}$ . If the spectrum of
$\hat{F}$ is bounded away from zero then $\hat{F}^{-1}$ is bounded
and one has the inclusions
\begin{equation}\label{sub}
  \mathfrak{H}_F\subset\mathfrak{H}\subset\mathfrak{H}_F^{-1}.
\end{equation}
\end{itemize}
We shall refer to the spaces $\mathfrak{H}_F$ and
$\mathfrak{H}_F^{-1}$ as a dual pair and when (\ref{sub}) is
satisfied, the three spaces $\mathfrak{H}_F$ ,$\mathfrak{H}$ and
$\mathfrak{H}_F^{-1}$ will be called a Gelfand triple
\cite{gel}.\\ \indent Let $\hat{B}$ be a (densely defined)
operator on $\mathfrak{H}$ and $\hat{B}^{\dag}$ its adjoint on
this Hilbert space. Assume that
$\mathcal{D}(B)\subset\mathcal{D}(F)$. Then unless
$[\hat{B},\hat{F}]=0$, the adjoint of $\hat{B}$, considered as an
operator on $\mathfrak{H}_F$ and which we denote by
$\hat{B}_F^{\ast}$ , is different from $\hat{B}^{\dag}$. Indeed,
\begin{eqnarray}
  \langle\psi|\hat{B}\phi\rangle_F&=&\langle\psi|\hat{F}\hat{B}\phi\rangle_\mathfrak{H}=\langle
  \hat{B}^{\dag}\hat{F}\psi|\phi\rangle_\mathfrak{H}=
  \langle \hat{F}\hat{F}^{-1}\hat{B}^{\dag}\hat{F}\psi|\phi\rangle_\mathfrak{H}\\ \nonumber &=& \langle
  \hat{F}^{-1}\hat{B}^{\dag}\hat{F}\psi|\phi\rangle_F.
\end{eqnarray}
Thus
\begin{equation}
  \hat{B}_F^{\ast}=\hat{F}^{-1}\hat{B}^{\dag}\hat{F}.
\end{equation}
Then due to the action of $\hat{T}$ on Hilbert space
$\mathfrak{H}$, we obtain other space $\mathfrak{H}_F$. Now if we
consider the oscillator operators $\hat{a}$, $\hat{a}^{\dag}$ and
$\hat{n}=\hat{a}^{\dag}\hat{a}$, we have the following operators
on $\mathfrak{H}_F$
\begin{equation}\label{aaa}
  \hat{A}_F=\hat{T}^{-1}\hat{a}\hat{T}\hspace{1cm}
  \hat{A}_F^{\dag}=\hat{T}^{-1}\hat{a}^{\dag}\hat{T}\hspace{1cm}\hat{n}_F=\hat{T}^{-1}\hat{n}\hat{T}.
\end{equation}
Clearly, considered as operators on $\mathfrak{H}_F$ , $\hat{A}_F$
and $\hat{A}^{\dag}_F$ are adjoints of each other and indeed they
are just the unitary transforms on $\mathfrak{H}_F$ of the
operators $\hat{a}$ and $\hat{a}^{\dag}$ on $\mathfrak{H}$. On the
other hand, if we take the operator $\hat{A}_F$ , let it act on
$\mathfrak{H}$ and look for its adjoint on $\mathfrak{H}$ under
this action, we obtain by (\ref{rrt}) the operator
$\hat{A}^{\sharp}=\hat{T}^{\ast}\hat{a}^{\dag}\hat{T}^{\ast-1}$
which, in general, is different from $\hat{A}^{\dag}_F$ and also
$[\hat{A}_F, \hat{A}^{\sharp}]\neq I$ , in general. In an
analogous manner, we shall define the corresponding operators
$\hat{a}_{F^{-1}}$, $\hat{a}^{\dag}_{F^{-1}}$, etc, on
$\mathfrak{H}_{F^{-1}}$. At this point we must mention, according
to this mathematical structure, operators $\hat{A}_F$ and
$\hat{A}^{\sharp}$ are exactly equivalent to generalized operators
defined in (\ref{defo}) that were adjoint of each other on the
same
Hilbert space $\mathfrak{H}$.\\
We use this mathematical structure to find proper representation
for the problem under consideration and by a constraint we will
determine operator $\hat{T}$.
\subsection{Quantization of fields}
In previous sections, we presented a description of the quantum
harmonic oscillator confined in a one dimensional infinite well
and we found its associated Heisenberg-type algebra. This algebra
is a deformed Heisenberg algebra which reduces to standard
Heisenberg algebra when the width of the well goes to infinity.\\
\indent
 Now using the hypothesis that
successive energy levels of the QHO confined in an infinite well
are obtained by creation or annihilation of quantum particles in a
box, we are going to construct a quantum field theory in a
confined region and using it to quantize EM field. We use
canonical field quantization approach. The Lagrangian associated
with a given field confined within a certain region can be
written as
\begin{equation}
\mathpzc{L}=\mathpzc{L}_{free}+V(r).
\end{equation}
where $\mathpzc{L}_{free}$ defines the Lagrangian of the free
field and $V(r)=\left\{\begin{array}{ll} 0 & \textrm{$-a\leq r\leq a$}\\
\infty & \textrm{elsewhere}
\end{array} \right.$. If we constrained the problem to the confined region $-a\leq
r\leq a$, the $V(r)=0$ and we have
$\mathpzc{L}=\mathpzc{L}_{free}$. This means that in the confined
region we can use the Lagrangian of the free field. Now if we
impose quantization postulate, this postulate will be the same as
free space. \\ \indent For example, we consider the EM field in a
confined region and in this region we have the following
Lagrangian for the field
\begin{equation}
\mathpzc{L}=-\frac{1}{4}F_{\mu\nu}F^{\mu\nu},
\end{equation}
where $F_{\mu\nu}=\partial_{\mu}A_{\nu}-\partial_{\nu}A_{\mu}$
($\mu,\nu=0,1,2,3$). As is customary in quantization of the EM
field we use the four-vector potential as the dynamical variable
of the field. We use the Coulomb Gauge in which
 $\vec{\nabla}\cdot\vec{A}=0$ and $A_0=0$.
 In this gauge, the Hamiltonian of the EM field is expressed in terms
 of the
vector potential $\vec{A}$ as \cite{field,QED}
\begin{equation}
H=\int d^3x\left[\left(\frac{\partial \vec{A}}{\partial t
}\right)^2+(\vec{\nabla}\times\vec{A})^2 \right].
\end{equation}
We consider the vector potential $\hat{\vec{A}}$ as the field
operator, and the quantization postulate for this field is
expressed by the following commutation relation (between
$\hat{\vec{A}}$ and its conjugate momentum,
$\vec{E}(r)=\frac{\partial\vec{A}}{\partial t}$)
\begin{equation}\label{quan}
[\hat{A}_i(\vec{r},t),\hat{E}_j(\vec{r'},t)]=-i\delta^3_{\perp
ij}(r-r'),
\end{equation}
where $\delta_{\perp}$ is the transverse delta function. Now, we
expand the field operator $\hat{\vec{A}}$ in terms of the ladder
operators of the confined QHO (from here we show creation and
annihilation operators of the confined QHO by $\hat{B}$ and
$\hat{B}^{\dag}$)
\begin{equation}
\hat{\vec{A}}=\sum_{\vec{k}}\frac{1}{\sqrt{2V\omega_{\vec{k}}}}\sum_{\lambda=1}^2\vec{\varepsilon}(\lambda,\vec{k})[\hat{B}_{k\lambda}
u_k(\vec r)+\hat{B}^{\dag}_{k\lambda}u_k^{\ast}(\vec r)],
\end{equation}
where $\vec{\varepsilon}$ is the polarization vector of the EM
field, $\lambda$ shows two independent polarization direction,
and $V$ is the volume of confinement. We interpret
$\hat{B}_{k\lambda}$ and $\hat{B}^{\dag}_{k\lambda}$,
respectively, as the annihilation and creation operators for a
deformed photon (quantum excitation of the confined EM field
under consideration) in direction $\vec{k}$, polarization
$\lambda$ and frequency $\omega_k$. The electric field operator
or the conjugate momentum associated with $\hat{\vec{A}}$ is
given by
\begin{equation}
\hat{\vec{E}}(r,t)=-\frac{\partial \hat{\vec{A}}}{\partial
t}=\sum_{\vec{k}}\frac{1}{\sqrt{2V\omega_k}}i\omega_k\sum_{\lambda=1}^2\vec{\varepsilon}(\lambda,\vec{k})
[\hat{B}_{k\lambda} u_k(\vec
r)-\hat{B}^{\dag}_{k\lambda}u_k^{\ast}(\vec r)].
\end{equation}
It is easy to show that
\begin{eqnarray}
[\hat{\vec{A}}_i(\vec
r,t),\hat{\vec{E}}_j(\vec{r'},t)]=&&\frac{-i}{2V}\sum_{\vec
k,\lambda}\varepsilon_i(\vec k,\lambda)\varepsilon_j(\vec
k,\lambda)\times\\ \nonumber &&[u_k(\vec
r)u_k^{\ast}(\vec{r'})+u_k^{\ast}(\vec
r)u_k(\vec{r'})]h(\hat{n}_{k,\lambda}),
\end{eqnarray}
where
$[\hat{B}_{k\lambda},\hat{B}^{\dag}_{k\lambda}]=h(\hat{n}_{k\lambda})=\gamma(2\hat{n}_{k\lambda}+1)+\alpha$.
 As is seen, in contrast to the quantization postulate (\ref{quan}), the right hand side of
the above commutator is an operator-valued function. Hence, if we
use the deformed operators $\hat{B}_{k\lambda}$,
$\hat{B}^{\dag}_{k\lambda}$ as amplitudes of the field expansion,
the quantization postulate imposed on the canonically conjugate
variables of the EM field is not preserved. To preserve the
commutation relation (\ref{quan}), we propose using another pair
of deformed operators in the Fourier decomposition of the field
operator.
 For this purpose, we consider the following dual operator of $\hat{B}$
\cite{dual}
\begin{equation}
\hat{B}=\hat{a}f(\hat{n})\hspace{1cm},\hspace{1cm}\hat{B}^{\dag}_f=\frac{1}{f(\vec{n})}\hat{a}^{\dag},
\end{equation}
which satisfy the commutation relation
\begin{equation}
[\hat{B}_{k\lambda},\hat{B}^{\dag}_{fk'\lambda'}]=\delta_{kk'}\delta_{\lambda\lambda'}.
\end{equation}
We use these operators to expand the field operator
\begin{equation}\label{field}
\hat{\vec{A}}=\sum_{\vec{k}}\frac{1}{\sqrt{2V\omega_k}}\sum_{\lambda=1}^2\vec{\varepsilon}(\lambda,\vec
k)[\hat{B}_{k\lambda} u_k(\vec
r)+\hat{B}^{\dag}_{fk\lambda}u_k^{\ast}(\vec r)].
\end{equation}
 As is clear, the operators $\hat{B}_{k\lambda}$ and
$\hat{B}^{\dag}_{fk\lambda}$ are not adjoint of each other with
respect to the ordinary scalar product, so the field operator is
not hermitian. It has been shown \cite{man3}, there is a
representation in which the operator $\hat{B}^{\dag}_f$ is
adjoint of the f-deformed operator $\hat{B}$ with respect to a
new scalar product in the carrier Hilbert space. Hence, in order
to preserve the quantization postulate, we should deform the
Hilbert space. We show the ordinary scalar product by
$\langle\,,\rangle$ and the deformed one by
$\langle\,,\rangle_f$. Since both scalar products are defined on
the same Hilbert space, they correspond to the same metric. The
relation between these two scalar product according to
(\ref{rrt}) can be written as
\begin{equation}\label{scal}
\langle\phi,\psi\rangle_f=\langle\phi,F\psi\rangle,
\end{equation}
where $F$ defines the relationship between two scalar products and
it can be determined from the condition that $\hat{B}$ and
$\hat{B}^{\dag}_f$ be adjoint of each other:
\begin{equation}
\langle\hat{B}\phi,\psi
\rangle_f=\langle\hat{B}\phi,F\psi\rangle=\langle\phi,\hat{B}^{\dag}F\psi\rangle
=\langle\phi,\hat{B}^{\dag}_f\psi\rangle_f.
\end{equation}
Therefore one can readily verify that $F$ is given by
\begin{equation}\label{nsa}
F=f^2(\hat{n})\prod_{m=1}^{\infty}f^2(\hat{n}-m).
\end{equation}
From Eqs.(\ref{rrt}) and (\ref{nsa}) operator $\hat{T}$ can be
found as
\begin{equation}
  \hat{T}=f(\hat{n})\prod_{m=1}^{\infty}f(\hat{n}-m).
\end{equation}
and according to Eq.(\ref{aaa}) the operators $\hat{B}_{k\lambda}$
and $\hat{B}^{\dag}_{k\lambda}$ can be obtained by the action of
$\hat{T}$. Now except other meaning of $T$ we can interpret it as
a transformation, that by its action ordinary system can be
changed to a confined system with definite barriers's position.\\
\indent Now instead of expanding the field operator in plane wave
basis we expand it in a basis that is orthogonal with respect to
the new scalar product (\ref{scal})
\begin{equation}\label{field1}
  \hat{\vec{A}}=\sum_{\vec{k}}\frac{1}{\sqrt{2V\omega_k}}\sum_{\lambda=1}^2\vec{\varepsilon}(\lambda,\vec
k)[\hat{B}_{k\lambda}v_k(\vec r
)+\hat{B}^{\dag}_{fk\lambda}v_k^{\ast}(\vec r)],
\end{equation}
where $v_k(\vec r )=\hat{T}u_k(\vec r)$, is a basis that is
orthogonal in the new representation as mentioned in mathematical
preliminary section. In this new representation the field
operator defined in Eq.(\ref{field1}) becomes Hermitian.
Furthermore, the electric field operator reads as
\begin{equation}
\hat{\vec{E}}(r,t)=\sum_{\vec{k}}\frac{i\omega_k}{\sqrt{2V\omega_k}}\sum_{\lambda=1}^2\vec{\varepsilon}(\lambda,\vec
k)[\hat{B}_{k\lambda} v_k(\vec
r)-\hat{B}^{\dag}_{fk\lambda}v_k^{\ast}(\vec r)],
\end{equation}
and the quantization postulate is recovered
\begin{equation}
[A_i(r,t),E_j(r',t)]=-i\delta^3_{\perp ij}(r-r').
\end{equation}
As mentioned before, in the confined region the Hamiltonian of
the EM field is the same as in free space. This Hamiltonian in the
Coulomb gauge is given by
\begin{equation}
\hat{H}=\frac{1}{2}\int
d^3r\left(\hat{\vec{E}}^2(r)+\hat{\vec{B}}^2(r)\right)=\frac{1}{2}\int
d^3r\left((\frac{\partial\hat{\vec{A}}}{\partial
t})^2+(\vec{\nabla}\times\hat{\vec{A}})^2\right),
\end{equation}
where $\hat{B}$ refer to the magnetic field. By substituting the
field operator $\hat{\vec{A}}$ given by (\ref{field1}) in the
above expression we arrive at the following Hamiltonian
\begin{equation}
\hat{H}=\sum_{k,\lambda}\omega_k\hat{B}^{\dag}_{fk\lambda}\hat{B}_{k\lambda}.
\end{equation}
Thus, the Hamiltonian can be interpreted as a collection of
f-oscillators for different modes of the EM field. The eigensates
of $\hat{H}$ which form a complete set and span the Hilbert space
of the system, are given by
\begin{equation}
|0\rangle ,\, \hat{B}^{\dag}_{fk\lambda}|0\rangle ,\,
\hat{B}^{\dag}_{fk\lambda}\hat{B}^{\dag}_{fk'\lambda'}|0\rangle,\cdots,
\end{equation}
where $|0\rangle$ is the vacuum state of the system i.e.
$\hat{B}_{k\lambda}|0\rangle=0$. In this manner, we interpret each
particle as an excitation of QHO confined in an infinite well.
This formulation can be used in confined systems and
nanostructures for considering elementary excitations, such as
ecxitons (which is a composite
excitation), phonons and plasmons.\\
\indent In quantum theory of fields, there are two important
concepts that are very useful in considering interacting fields.
One of them is Feynman propagator which is defined for a general
field operator $\hat{\psi}(x)$ as \cite{field}
\begin{equation}
iD_{F}(x-y)=\langle0|\hat{C}(\hat{\psi}(x)\hat{\psi}(y))|0\rangle,
\end{equation}
where $\hat{C}$ is the time-ordered operator (we show the time
ordering operator by $\hat{C}$ for making distinction between
this operator and the operator $\hat{T}$ defined in
(\ref{hehehe})). Now, if we assume that the field under
consideration is spatially confined, then according to the
definition of the deformed scalar product given by (\ref{scal})
the corresponding Feynman propagator is defined as
\begin{equation}
iD'_{F}(x-y)=_f\langle0|\hat{C}(\hat{\psi}(x)\hat{\psi}(y))|0\rangle_f.
\end{equation}
Making use of this definition for the photon field in a confined
region and applying field operator (\ref{field1}) result in:
\begin{equation}\label{fey}
D'_F(x-y)=F(0)D_F(x-y).
\end{equation}
where
$F(\hat{n})=f^2(\hat{n})\prod_{m=1}^{\infty}f^2({\hat{n}-m})$.
Eq.(\ref{fey}) shows that the Feynman propagator has not any
difference in confined field theory except a constant factor that
depends on some physical parameters such as the size of the
system, and reduces to the standard propagator when the
boundaries tend to infinity. Another important concept is the
scattering matrix (S matrix), that describes the probability
amplitude for a process in which the system makes a transition
from an initial state to a final state under the influence of an
interaction. According to the concept of S matrix, the
probability amplitude for a transition from the initial state
$|i\rangle$ into the final state $|f\rangle$ is defined as
\begin{equation}
 S_{fi}=\langle f|\hat{S}|i\rangle,
\end{equation}
where operator $\hat{S}$ is defined in terms of the interaction
Hamiltonian in the interaction picture as \cite{field}
\begin{equation}
\hat{S}=\sum_{l=0}^{\infty}\frac{1}{l!}(-i)^l\int d^3r_1\cdots
d^3r_l\hat{C}[H_{int}(t_1)\cdots H_{int}(t_l)].
\end{equation}
In our formalism, according to the new definition of scalar
product we define the probability amplitude as
\begin{equation}
S'_{fi}=\langle f|\hat{S}|i\rangle_f=\langle
f|F(\hat{n})\hat{S}|0\rangle.
\end{equation}
and due to the concept of Fock states we have
\begin{equation}
S'_{fi}=F(n)\langle f|\hat{S}|i\rangle=F(n)S_{fi}.
\end{equation}
This equation shows that the new S matrix is proportional to the
standard S matrix with a constant of proportionality that is a
function of number of quantum excitations. Furthermore, we can
conclude that spatial confinement of an interacting system
results in an intensity-dependent coupling constant. As an
example, consider an EM field that interacts with a fermionic
system in a confined region. We assume that in this system,
fermions be expressed by undeformed Dirac field operator denoted
by $\hat{\psi}$ , and photons are described by (\ref{field1}).
The interaction Hamiltonian can be written as
\begin{equation}
\hat{H}_{int}=-e\hat{\overline{\psi}}\gamma\hat{\psi}\vec{A}.
\end{equation}
Therefore the $S$ matrix is given by
\begin{eqnarray}\label{smat}
S_{fi}&=&1-ieF(n)\int
d^3x:\hat{\overline{\psi}}(x)\gamma\hat{\psi}(x)\vec{A}(x):\nonumber
\\ &+&\frac{(-ie)^2F(n)}{2!}\int
d^3xd^3y\hat{C}\left[:\hat{\overline{\psi}}(x)\gamma\hat{\psi}(x)\vec{A}(x):
:\hat{\overline{\psi}}(x)\gamma\hat{\psi}(x)\vec{A}(x):\right]+\cdots\\
\nonumber
\end{eqnarray}
where the symbol $::$ denotes the normal ordering. So one can
conclude from (\ref{smat}) that coupling constant in each term of
the expansion has a same dependence on the intensity of the photon
field. Dependence of coupling constant on intensity is a
indication of nonlinear interaction.
\section{Generation of coherent states in a confined region}
In this section we consider an infinite well directed in the
z-direction, in which we have a current density ($A_0\neq0$). For
example, an electron that moving in axial direction of the well
generates the following classical current
\begin{equation}
\vec{j}=ev\delta(x)\delta(y)\delta(z-vt)\hat{k},
\end{equation}
where $v$ is the velocity of electron. In the presence of current
density, the equation of motion for the vector potential $\vec{A}$
 (according to the Maxwell equations) reads as
\begin{equation}
\frac{1}{c^2}\frac{\partial^2\vec{A}}{\partial
t^2}+\vec{\nabla}\times\vec{\nabla}\times\vec{A}=\frac{1}{c}\vec{j'},
\end{equation}
where $\vec{j'}=\vec{j}-\vec{\nabla}\varphi$ is the transverse
part of the current density. This equation can be derived from the
Hamiltonian
\begin{eqnarray}\label{ham}
\hat{H}&=&\int
d^3r\left[\frac{1}{2}\left(\left(\frac{\partial\hat{\vec{A}}}{\partial
t}\right)^2+\left(\vec{\nabla}\times\hat{\vec{A}}
\right)^2\right)+\frac{1}{c}\vec{j'}(r,t)\centerdot\hat{\vec{A}}(r,t)\right] \nonumber \\
&=&\sum_{k,\lambda}\left(\omega_k\hat{B}^{\dag}_{fk\lambda}\hat{B}_{k\lambda}\right)+\frac{1}{c}\int
d^3r\vec{j'}(r,t)\centerdot\hat{\vec{A}}(r,t) \nonumber\\
&=&\sum_{k,\lambda}[\omega_k\hat{B}^{\dag}_{fk\lambda}\hat{B}_{k\lambda}
\nonumber \\
 &+&\frac{1}{c}\frac{\vec{\varepsilon}(k,\lambda)}
{\sqrt{2V \omega_k}}\int
d^3r\left(\hat{B}_{k\lambda}g(k,r)+\hat{B}^{\dag}_{fk\lambda}g^{\ast}(k,r)\right)\centerdot\vec{j'}(r,t)],
\end{eqnarray}
The transverse density current $\vec{j}'$ and the polarization
vectors are in the same plane and are in the same direction. The
Hamiltonian (\ref{ham}) can be rewritten as
\begin{equation}
\hat{H}=\sum_{k,\lambda}\left[\omega_k\hat{B}^{\dag}_{fk\lambda}\hat{B}_{k\lambda}+\frac{1}{\sqrt{2V\omega_k}}
\left(\hat{B}_{k\lambda}j'(k,t)+\hat{B}^{\dag}_{fk\lambda}j'^{\ast}(k,t)\right)\right].
\end{equation}
where $j'(k,t)=\frac{1}{c}\int
d^3r\vec{\varepsilon}(k,\lambda)\cdot\vec{j}'(r,t)g(k,r)$. The
equation of motion for $\hat{B}_{k\lambda}(t)$ that follows from
the above Hamitonian reads as
\begin{equation}\label{equm}
\dot{\hat{B}}=-i\omega_k\hat{B}_{k\lambda}-i\frac{j'^{\ast}(k,t)}{\sqrt{2V\omega_k}}.
\end{equation}
If we define a new variable
$\widetilde{\hat{B}}_{k\lambda}(t)=e^{i\omega_kt}\hat{B}_{k\lambda}$,
the solution of Eq.(\ref{equm}) is
\begin{equation}
\widetilde{\hat{B}}_{k\lambda}(t)=\widetilde{\hat{B}}_{k\lambda}(-\infty)-\frac{i}{\sqrt{2V\omega_k}}\int_{-\infty}^t
j'^{\ast}(k,t')e^{i\omega_kt'}dt'.
\end{equation}
The time dependence of the operator
$\widetilde{\hat{B}}_{k\lambda}(t)$ can be regarded as a result of
the following unitary transformation
\begin{eqnarray}
\widetilde{\hat{B}}_{k\lambda}(t)&=&\hat{O}^{\dag}\hat{B}_{k\lambda}(-\infty)\hat{O},\nonumber
\\
\hat{O}&=&\exp\left[\sum_{k,\lambda}\left(\alpha(k,t)\hat{B}^{\dag}_{fk\lambda}(-\infty)-
\alpha^{\ast}(k,\lambda)\hat{B}_{k\lambda}(-\infty)\right)\right].
\end{eqnarray}
where by definition
$\alpha(k,t)=\frac{-i}{\sqrt{2V\omega_k}}\int_{-\infty}^t
j'^{\ast}(k,t')e^{i\omega_kt'}dt'$. The operator $\hat{O}$ is a
displacement-like operator \cite{dual,man3}. If we choose the
initial state of the EM field to be the vacuum $|0\rangle$, then
the state vector at time $t$ is
\begin{eqnarray}
|\beta(k,t)\rangle&=&\hat{O}|0\rangle=\exp\left[\sum_{k,\lambda}\left(\beta(k,t)\hat{B}^{\dag}_{fk\lambda}(-\infty)-
\beta^{\ast}(k,\lambda)\hat{B}_{k\lambda}(-\infty)\right)\right]|0\rangle\nonumber\\
&=&e^{-\frac{|\beta(k,t)|^2}{2}}\sum_n
\frac{\beta^n(k,t)}{[f(n)]!n!}(\hat{a}^{\dag})^n|0\rangle
\nonumber \\
&=&e^{-\frac{|\beta(k,t)|^2}{2}}\sum_n\frac{\beta^n(k,t)}{\sqrt{n!}[f(n)]!}|n\rangle.
\end{eqnarray}
In the sense of Eqs.(\ref{coh})-(\ref{coh1}) it is evident that
this state can be regarded as a nonlinear coherent state.
\section{Conclusion}
In this paper, we have considered the relation between the
spatial confinement effects and special kind of f-deformed
algebra. We have found that the confined simple harmonic
oscillator can be interpreted as an f-oscillator, and we have
obtained the corresponding deformation function. Then we have
searched the effects of boundary conditions in quantum field
theory. We have used f-deformed operators as the dynamical
variables and found that for preserving commutation relation
between the field operator and its conjugate momentum we should
deform Hilbert space of the system under consideration. As a
result of new definition of scalar product, we have concluded
that the coupling constant of interactions in confined systems
become a function of number of excitation, for example in the
case of EM field coupling constant becomes a function of intensity
of EM field. Finally we have proposed a theoretical scheme for
generating nonlinear coherent states of EM field through the
coupling of a classical current to the vector potential operator
$\hat{\vec{A}}$ inside a confined region.

{}
\newpage

\section*{References}
\begin{thebibliography}{}
\bibliographystyle{amsplain}
\bibitem{mesos}D. K. Ferry, S. M. Goodnick, \textit{Transport in
nanostructures}, (Cambridge University Press 1999).\\
\hspace{0.7cm} B. L. Altshuler, P. A. Lee, R. A. Web,
\textit{Mesoscopic phenomena in solids}, (North-Holland,
Amsterdam 1991).
\bibitem{casimir}G. Plunien, B. M\"{u}ller, W. Greiner, Phys. Rep.
\textbf{134}, 87 (1986).
\bibitem{theory}H. Haug, S. W. Koch, \textit{Quantum theory of the optical and electronic properties
of semiconductors}, 4th ed. (World Scientific, Singapore, 2004).
\bibitem{expri}P. Michler, \textit{Single quantum dots: Fundamental, Application, and new
concepts}, Topics in applied physics (Springer, 2003).
\bibitem{1}P. Michler, A. Imamo\u{g}lu, M. D. Mason, P. J.
Carson, G. F. Strouse, S. K. Buratto, Nature
(London)\textbf{406}, 968 (2000);\\
P. Michler, A. Kiraz, C. Becher, W. V. Shoenfeld, P. M. Petroff,
L. Zhang, E. Hu, A. Imamo\u{g}lu, Science \textbf{290}, 2282
(2000).
\bibitem{2}P. Michler, A. Kiraz, L. Zhang, C. Becher, E. Hu, A.
Imamo\u{g}lu, Appl. Phys. Lett. \textbf{77}, 184 (2000).
\bibitem{3}T. Feldtmann, L. Schneebeli, M. Kira, S. W. Koch, Phys.
Rev. B \textbf{73}, 155319 (2006);\\
C. Gies, J. Wiersing, M. Lorke, F. Jahnke, Phys. Rev A
\textbf{75}, 013803 (2007);\\
K. Ahn, J. F\"{o}rstner, A. Knorr, Phys. Rev. B \textbf{71},
153309 (2005).
\bibitem{field}W. Greiner, J. Reihardt, \textit{Field
quantization} (Springer-Verlag 1996).\\
\hspace{0.7cm}Steven Weinberg, \textit{The quantum theory of
fields Vol.1} (Cambridge University Press 1995).
\bibitem{man1}V. I. Man'ko, G. Marmo, G. C. E. Sudarshan, F.
Zaccaria, Phys. Scr. \textbf{55}, 528 (1997).
\bibitem{man2}V. I. Man'ko, R. Vilela Mendes, J. Phys. A: Math.
Gen.
\textbf{31}, 6037 (1998);\\
\hspace{0.7cm}V. I. Man'ko, G. Marmo, S. Solimeno, F. Zaccaria,
Int. J. Mod. Phys. A \textbf{8}, 3577 (1993);\\
V. I. Man'ko, G. Marmo, S. Solimeno, F. Zaccaria, Phys. Lett. A
\textbf{176}, 173 (1993).
\bibitem{hame}V. I. Man'ko, G. Marmo, E. C. G. Sudarshan, and F.
Zaccaria, in: N. M. Atakishiev (Ed), Proc. IV Wigner Symp.
(Guadalajara, Mexico, July 1995), World Scientific, Singapore,
1996, P.421; S. Mancini, Phys. Lett. A \textbf{233}, 291, 55
(1997); S. Sivakumar, J. Phys. A: Math. Gen. \textbf{33}, 2289
(2000); B. Roy, Phys. Lett. A \textbf{249}, 55 (1998); H. C. Fu,
R. Sasaki, J. Phys. A: Math. Gen. \textbf{29}, 5637 (1996); R.
Roknizadeh, M. K. Tavassoly, J. Phys. A: Math. Gen. \textbf{37},
5649 (2004); M. H. Naderi, M. Soltanolkotabi, R. Roknizadeh, J.
Phys. A: Math. Gen. \textbf{37}, 3225 (2004); A-S. F. obada, G.
M. Abd Al-Kader, J. Opt. B: Quantum Semiclass. Opt. \textbf{7},
S635 (2005).
\bibitem{vogel}R. L. deMatos Filho, W. Vogel, Phys. Rev. A
\textbf{54}, 4560 (1996).
\bibitem{naderi}M. H. Naderi, M. Soltanolkotabi, R. Roknizadeh,
Eur. Phys. J. D \textbf{32}, 397 (2005).
\bibitem{nondet}M. H. Naderi, PhD Thesis, University of Isfahan,
Iran (unpublished 2004).
\bibitem{mahdi}A. Mahdifar, R. Roknizadeh, M. H. Naderi, J. Phys.
A: Math. Gen. \textbf{39}, 7003 (2006).
\bibitem{bezze}V. B. Bezzera, M. A. Rego-Monteiro, Phys. Rev. D
\textbf{70}, 065018 (2004).
\bibitem{cur}E. M. F. Curado, M. A. Rego-Monteiro, J. Phys. A:
Math. Gen. \textbf{34}, 3253 (2001).
\bibitem{swamy}P. Narayana Swamy, Physica A \textbf{353}, 119
(2005).
\bibitem{naderi1}M. H. Naderi, M. Soltanolkotabi, R. Roknizadeh,
J. Phys. Soc. Japan \textbf{73}, 2413 (2004).
\bibitem{hame1}S. Sivakumar, Phys, Lett. A \textbf{250}, 257
(1998); V. I. Man'ko, A. W\"{u}nshe, Quantum Semiclass. Opt.
\textbf{9}, 381 (1997); G. N. Jones, J. Haight, C. T. Lee,
Quantum Semiclass. Opt. \textbf{9}, 411 (1997); V. V. Dodonov, Y.
A. Korennoy, V. I. Man'ko, Y. A. moukhin, Quantum Semiclass. Opt.
\textbf{8}, 413 (1996); Nai-li Liu, Zhi-hu Sun, Hong-yi Fan, J.
phys. A: Math. Gen. \textbf{33}, 1993 (2000);B. Roy, P. Roy, J.
Opt. B: Quantum Semiclass. Opt. \textbf{1}, 341 (1999).
\bibitem{CQHO}G. Campoy, N. Aquino, V. D. Granados, J. Phys. A:
Math. Gen.
\textbf{35}, 4903 (2002);\\
\bibitem{agu}V. C. Aguilera-Navarros, E. Ley Koo, A. H. Zimerman, J. Phys.
A: Math. Gen. \textbf{13}, 3585 (1980).
\bibitem{zico}C. Zicovich-Wilson, J. H. Planelles, W. Jackolski,
Int. J. Quan. Chem. \textbf{50}, 429 (1994).
\bibitem{fac}L. Infeld, T. E. Hull, Rev. Mod. Phys. \textbf{23},
21 (1951).
 \bibitem{klauder}J. R. Klauder, B. Skagerstam, \textit{Coherent states: Application in Physics and Mathematical
 Physics} (World Scientific, Singapore, 1985).\\ \hspace{0.5cm} S.
 T. Ali, J-P. Antoine, J-P. Gazeau, \textit{Coherent states, Wavalets, and their generalization
  (New York: Springer 2000)}
 \bibitem{bessel}G. N. Watson, \textit{Theory of Bessel
 functions}, Second ed. (Cambridge University Press 1966) P.388.
 \bibitem{mandel} L. Mandel, E. Wolf, \textit{Optical coherence and quantum
optics} (Cambridge University Press 1995).
\bibitem{tavasol}S. Twareque Ali, R. Roknizadeh, M. K. Tavassoly,
J. Phys. A: Math. Gen. \textbf{37}, 4407 (2004).
\bibitem{gel}M. I. Gelfand, N. Ya Vilenkin, \textit{Generalized
functions} Vol. 4 (New York: Academic Press 1964).
\bibitem{QED}W. Hitler, \textit{The quantum theory of radiation}, Third Edition (Clarendo Press 1954).
\bibitem{dual}B. Roy, Phys. Lett. A \textbf{249}, 25 (1998);\\
B. Roy, P. Roy, Phys. Lett. A \textbf{263}, 48 (1999).
\bibitem{man3}P. Aniello, V. Man'ko, G. Marmo, S. Solimeno, F.
Zaccaria, J. Opt. B: Quantum Semiclass. Opt. \textbf{2}, 718
(2000).
\end{thebibliography}
\end{document}